\documentclass{amsart}

\usepackage{graphicx}
\usepackage{amssymb}
\usepackage{amsthm}

  \theoremstyle{definition} \newtheorem {definition} {Definition}

  \newcommand{\ccn}[1]{\mathsf{{#1}}}

\begin{document}

\title{Complexity-Style Resources in Cryptography}

\author{Ed Blakey}
\address{School of Mathematics, University of Bristol, University Walk, Bristol, BS8 1TW, UK}
\email{ed.blakey@queens.oxon.org}

\begin{abstract}
Previously, the author has developed a framework within which to quantify and compare the resources consumed during computational---especially \emph{unconventional} computational---processes (adding to the familiar resources of \emph{run-time} and \emph{memory space} such non-standard resources as \emph{precision} and \emph{energy}); it is natural and beneficial in this framework to employ various complexity-theoretic tools and techniques.
Here, we seek an analogous treatment not of computational processes but of \emph{cryptographic protocols} and similar, so as to be able to apply the existing arsenal of complexity-theoretic methods in new ways, in the derivation and verification of protocols in a wider, cryptographic context.
Accordingly, we advocate a framework in which one may view as resources the costs---which may be related to computation, communication, information (including side-channel information) or availability of primitives, for example---incurred when executing cryptographic protocols, coin-tossing schemes, etc.
The ultimate aim is to formulate as a resource, and be able to analyse complexity-theoretically, the \emph{security} of these protocols and schemes.
\end{abstract}


\maketitle

\section{Introduction}

We begin by outlining in Sect.~\ref{sec:intbac} the notion of \emph{resource} as already used in a computational context, and by motivating in Sect.~\ref{sec:intmot} extension of the notion to a cryptographic setting.
This leads us to advocate in Sect.~\ref{sec:crs} a resource-centric framework in which to derive and analyse cryptographic protocols.

\subsection{Background}\label{sec:intbac}

\subsubsection{Computation}

For present purposes, one may view computation as the conversion by some machine, device or system (the \emph{computer}) of an \emph{input value} into an \emph{output value}.
The details of the computer and its implementation, and even of the model of computation (e.g.,\ Turing machine, analogue system or quantum computer) to which it conforms, are unimportant; sufficient is that the computer have provision for accepting an input value, performing some form of processing, and supplying an output value.
These three stages together can be viewed as the evaluation of a function\footnote{To some input values there may in fact correspond several potential output values, whence not a function but rather a \emph{multifunction} is computed.} (that which is computed) at a given value (the input value) of the function's argument.

\subsubsection{Complexity}\label{sec:intbaccpx}

Of course, for a computer to be of practical use, it should be in some sense efficient.
Receiving from the computer a correct output value is not in itself sufficient if, for example, the value is not supplied in reasonable \emph{time}: a computation that, between input and output, sees the age of the universe elapse is of no real-world use.

One formalizes such ideas of efficiency using \emph{computational complexity theory}, wherein the resources consumed during, or required for, computation are quantified.
The most common form of complexity analysis is measurement of the \emph{time} that elapses during a computation.
In the case of a Turing machine, for example, this quantity is the number of time-steps that elapse; in that of a physical system such as an analogue or optical computer, it is the physical time (in seconds, say) between provision of an input value and receipt of an output value.

Alongside time, one may measure the consumption during computation of \emph{space}.
For a Turing machine, this is defined to be the number of (distinct) tape-cells to which the machine writes during the computation; for a physical (e.g.,\ analogue or optical) system, this is rather the physical volume\footnote{More precisely, one measures the minimal volume of a cuboid bounding the apparatus; this is a more natural and insightful choice than the volume of the computer itself if, for example, the form of the computer resembles a space-filling curve, occupying little actual volume but nonetheless requiring a large contiguous space.} occupied by the apparatus (including any storage space that may be required during computation).

Time and space, then, are examples of \emph{computational resources}.
These are commodities consumed in some quantity during the computation, or otherwise required in some quantity for the computation to succeed.
We formalize these as functions that map input values to the corresponding required amounts of resource, which we take to be natural numbers, rounding or similar when necessary; for example, a Turing machine $M$ has \emph{time} and \emph{space} resource functions $T_M$ and $S_M$ respectively (or simply $T$ and $S$ when $M$ is understood) where $T\left(x\right)$ is the number of time-steps that elapse, and $S\left(x\right)$ the number of tape-cells to which are written, as $M$ computes given input $x$ (in fact, when the computers under consideration are Turing machines, time and space are up to variation the \emph{only} computational resources; unconventional, non-Turing systems, on the other hand, may well consume unconventional resources, as can be seen below).

As is to be expected, such quantities of required resource often depend upon the size\footnote{The appropriate notion of size here depends upon context. A very frequently encountered example sees the input value take the form of a natural number $k$ presented in binary notation, in which case the appropriate definition of the size $\left\vert k\right\vert$ is the number of bits in $k$, or more typically the approximation $\log_2\left(k\right)$ thereof. See for example \cite{papa} for further discussion.} of the input value from which the computation begins: if one wishes to compute with larger and larger values, then it may well be that the system will require more and more time to perform the computation, and more and more memory space in which to store the intermediate values with which it works.
From computational resources, then, one may derive corresponding \emph{complexity functions}---these are resource consumption viewed as functions of input size.
Broadly, whereas a resource function $A$ maps input values $x$ to the corresponding required, natural-number amounts of resource, the associated \emph{complexity function} $A^*$ maps input \emph{sizes} $n$ to the corresponding \emph{maximum} required amounts of resource:
\begin{equation}\label{eq:cpxfn}
A^*\left(n\right) := \sup\left\{\, A\left(x\right) \;\middle\vert\; \left\vert x\right\vert = n \,\right\} \enspace .
\end{equation}
These are the functions that are spoken of in complexity theory as being logarithmic, polynomial, exponential, etc.,\ and they indicate something of the efficiency of computing systems,\footnote{As a rule of thumb, \emph{efficiency} is identified with resource consumption that grows no more quickly than \emph{polynomially} in the size of the input value.} which in turn indicates something of the difficulty of the problems that the systems solve.

We mention above the standard resources of \emph{space} and \emph{time}.
We claim now (and defer to \cite{thesis} detail and discussion beyond the comments of the following paragraphs) that, \emph{when dealing with `standard', digital computers} (Turing machines, for example), consideration of no other resources than these two is necessary.
Indeed, the complexity classes of standard complexity theory---some of the more commonly encountered being $\ccn{P}$, $\ccn{NP}$, $\ccn{coNP}$, $\ccn{PH}$, $\ccn{PSPACE}$, $\ccn{EXP}$, $\ccn{AC^0}$, $\ccn{NC}$, $\ccn{L}$, $\ccn{P/poly}$, $\ccn{BPP}$, $\ccn{BQP}$ and $\ccn{PP}$---are defined in terms of time and/or space, but of no other resources.\footnote{The classes listed here are all in the `Petting Zoo' section of Scott Aaronson's Complexity Zoo \cite{zoo}, and as such are amongst purportedly the most important (the most referenced or fundamental, say) classes. Neither are the classes that are in the Petting Zoo but not listed here defined in terms of resources other than time and space.}

One may argue, contrarily to this claim, that certain of these standard complexity classes \emph{are} in fact defined in terms of resources other than time and space.
For example, $\ccn{NP}$ is defined with reference to \emph{non-determinism}, and $\ccn{NC}$ with reference to \emph{parallelism}; non-determinism and parallelism, it can be argued, are resources in the sense that their use apparently confers computational advantage.
However, they are not resources in the same sense as, say, time and space: they are not \emph{commodity} resources (see below and \cite{ucuc}).
Each of non-determinism and parallelism is either permitted by a given computational model, or it is not; whether it is depends solely upon the choice of computational model, and in particular not upon the choice of input value, and so no complexity functions associated with these two `resources' arise in the same way (i.e.,\ as defined by (\ref{eq:cpxfn})) as that in which time and space complexity arise.

The \emph{commodity} resources (e.g.,\ time but not non-determinism) on which we focus in Sect.~\ref{sec:intbac} are exactly the resources suitable for specifying the boundaries of complexity classes: whereas $\ccn{NP}$ is defined with reference to non-determinism, for example, this is only in the capacity of establishing what is admitted as a computer (namely, a non-deterministic Turing machine), rather than how much of some commodity the computer is allowed to consume during computation; time, on the other hand, \emph{bounds} $\ccn{NP}$---one is allowed polynomial time and no more\mbox{---,} and thus describes the class's frontier.

We reiterate that, when one considers only standard computers, there are but two applicable commodity resources: time and space.
However, there exist also non-standard computers---quantum, chemical and optical systems, for example\mbox{---,} which may well consume correspondingly non-standard resources.
We mention now an important such resource, \emph{precision}, from the author's previous work (see, for example, \cite{thesis} for detail further than that given below).

\subsubsection{Precision}\label{sec:intbacpre}

Standard devices such as Turing machines, finite-state automata and real-life digital computers operate with discrete (input, intermediate and output) values such as natural numbers or bits.

The very definition \cite{turing} of a Turing machine gives that one can \emph{discern} which symbol is stored at a given tape location, and which state the machine occupies: the Turing machine never presents the difficulty of having to resolve, say, an ambiguously written symbol that could be either a `$0$' or a `$1$'.
That this is so arises from the formalization of the respective collections of states and symbols as (finite) \emph{sets}, which do not, of course, admit repetition; the states, and similarly the symbols, are mathematically distinct elements, with no structure thereamongst based upon proximity or similarity.

Furthermore, that the values accepted, processed and supplied by a computing system are discrete is a property satisfied not only by the abstract Turing machine, but also by the real-world digital computer.
Physical implementation sees the abstract $0$s and $1$s realized, for example, as potential differences of $0$ and $5\,\mathrm{volts}$ respectively; thus, the design of digital computers sacrifices the possibility of simultaneously conveying many bits over a single physical connection (e.g.,\ by utilizing potential differences taken from a \emph{continuous} range), but in so sacrificing maintains easily distinguishable values: the relatively undemanding requirement of being able to discern potential difference to within $2{\cdot}5\,\mathrm{volts}$ guarantees that values of $0$ and $5\,\mathrm{volts}$ are distinguishable, whence the intended bit is correctly retrieved.

However, this discrete nature is not shared by all computers.
We give three examples.
\begin{itemize}
  \item Consider first an \emph{optical computer}, to which is conveyed an input value encoded in the wavelength of a light source (cf.\ the systems of \cite{facana} and Sect.~\ref{sec:intbacafs} below); the user's means of supplying to the system the input value is, for example, to manipulate a variable resistor that controls the availability of energy to the source, and consequently the wavelength of the light produced thereby.
However, whereas the user has in mind a specific input value $x$ with which he wishes to compute, his ability physically to manipulate the resistor may well be marred by imprecision; as a result, the value actually received by the system may be not $x$, but rather some arbitrary element of $\left[x - \epsilon, x + \epsilon\right]$.
Similarly, it may be that the \emph{output} value is presented by the system encoded in, say, the distance between two points of light on a screen; if the user has manually to measure this distance, then imprecision will once again corrupt the true value.
  \item A similar situation is encountered with \emph{analogue computers} (see \cite{bush} for a notable example, and \cite{thesis} for associated precision-related discussion): if the user is required physically to manipulate parameters of the computer (the angle of shafts, for example) so as to effect input, and physically to measure parameters of the system (dial readings, for example) so as to effect output, then imprecision may be introduced as in the preceding example.
  \item A further example with broadly the same features concerns the \emph{slide rule}.
The user will be able with arbitrary precision neither to line up the two scales of the rule nor to read the value that becomes lined up with a certain notch.
\end{itemize}
Of course, the pattern common to these examples is that the computations' input and output values are encoded in the values of physical parameters, such as wavelength, distance and angle.
The sets of possible values for these parameters are \emph{continua} (specifically, intervals of real numbers),\footnote{This is true, at least, under the assumption that the computers operate at a super-quantum scale. In cases where this assumption fails, and notably when quantum phenomena are actively exploited so as to \emph{aid} computation, one may encounter different forms of imprecision. It remains, however, that precision, whatever exact form it may take, is a computational resource relevant to these physical computations.} and the physical manipulation and measurement of these parameters will be prone to error.

All is not lost.
Sufficiently small imprecisions may be corrigible, and one may in fact \emph{quantify} the precision required (in order that resultant errors can be corrected) of the user during input and output---we describe this quantification below.
This gives rise to a computational resource, \emph{precision}---a commodity resource in the same sense as time and space (recall Sect.~\ref{sec:intbaccpx}).

Our notion of precision deals specifically with the input and output processes of a computation (in particular, precision captures the computer's lack of robustness against input/output imprecision).
We formalize the relevant aspects of computation as follows (and defer further detail to \cite{thesis}).
\begin{itemize}
  \item The computer has a number of physical \emph{parameters} that effect input and output: the user conveys to the system his intended input value by manipulating the \emph{input parameters}, and, after the computation has taken place, receives the corresponding output value by measuring the \emph{output parameters}.
Each of the $p \in \mathbb{N}$ input parameters is modelled as a pair $\left(i_j, V_{i_j}\right)$ ($1 \leq j \leq p$), where $i_j$ is an \emph{input} and $V_{i_j}$ the set of \emph{values} to which $i_j$ may be set; each output parameter is a pair $\left(o_k, V_{o_k}\right)$ ($1 \leq k \leq q \in \mathbb{N}$), where $o_k$ is an \emph{output} and $V_{o_k}$ is the set of \emph{values} that $o_k$ may take.
  \item Consequently, an \emph{input value} is an assignment $x \in V_{i_1} \times {\ldots} \times V_{i_p}$ of valid values to inputs (where input $i_j$ is considered to be assigned the value $\pi_j\left(x\right) \in V_{i_j}$); an \emph{output value}, similarly, is an assignment $y \in V_{o_1} \times {\ldots} \times V_{o_q}$ of valid values to outputs (where output $o_k$ is considered as having taken the value $\pi_k\left(y\right) \in V_{o_k}$).
  \item The \emph{computation relation} (denoted `$\smile_\Phi$', where $\Phi$ is the computer in question) relates an input value to all corresponding output values that can result (those of which the user seeks one)---this relation is the multifunction computed by $\Phi$.
A deterministic computation, then, has as its computation relation a (partial) \emph{function} that maps each input to its unique output (if defined), whereas non-deterministic computations give rise to more general relations.
  \item There are two important input values that, due to imprecision in the physical process of the user's manipulating the input parameters, may well differ and that, accordingly, we distinguish: the input value \emph{intended} by the user, and that actually \emph{implemented} by the user.
Though imprecise adjustment of the input parameters may well render these values unequal, they are, nonetheless, mutually constrained by a relation that depends upon the details of implementation of the input parameters.
For example, it may be that the user wishes to adjust a dial (input $i$, say) to an angle $\theta \in V_{i}$, but is able to guarantee of the angle $\theta'$ \emph{actually} implemented only that $\left|\theta - \theta'\right| \leq \epsilon_{i}$, where non-negative, real number $\epsilon_{i}$ depends upon the precise implementation details of $i$ and reflects the fidelity with which this input accepts its value.\footnote{That $\theta' \in \left[\theta - \epsilon_{i}, \theta + \epsilon_{i}\right]$ ($\epsilon_{i} \geq 0$) follows from our implicitly supposing an \emph{additive} error in using $i$; the error may instead be \emph{multiplicative}---in which case we should have rather that $\theta' \in \left[\theta / \epsilon_{i}, \epsilon_{i}\theta\right]$ ($\epsilon_{i} \geq 1$)\mbox{---,} or may obey some other relation (nonetheless characterized by real-number error term $\epsilon_{i}$) constraining $\theta$ and $\theta'$.}
Whilst the intended and implemented input values typically differ, then, they are at least related by the \emph{input error relation} $R_{\epsilon_I}$, characterized by the tuple $\epsilon_I := \left(\epsilon_{i_1}, {\ldots}, \epsilon_{i_p}\right) \in \mathbb{R}^p$ of individual input parameters' error terms (which, in turn, depend upon the details of the process whereby the user adjusts these parameters): an attempt by the user to set the input parameters to the (intended) input value $x$ results with strictly positive probability in the parameters' actually receiving (implemented) input value $x'$ if and only if $x \, R_{\epsilon_I} \, x'$.
  \item Similarly, there may well be inequality between two important output values: the \emph{true} output value, to which the output parameters are set (during and by the computation), and the \emph{measured} output value, which is the outcome of the user's (typically imprecisely) measuring the true output value.
Each output $o_k$ (more specifically, the imprecision arising from measurement of the output's value) is characterized by non-negative, real error term $\epsilon_{o_k}$, just as $\epsilon_{i_j}$ characterizes $i_j$ in the previous bullet point.
The true and measured output values, though quite possibly different, are nonetheless related by the \emph{output error relation} $R_{\epsilon_O}$, characterized by the tuple $\epsilon_O := \left(\epsilon_{o_1}, {\ldots}, \epsilon_{o_q}\right) \in \mathbb{R}^q$ of individual output parameters' error terms (which, in turn, depend upon the details of the process whereby the user measures these parameters): an attempt by the user to measure output parameters set to the (true) output value $y'$ results with strictly positive probability in the (measured) output value $y''$ if and only if $y' \, R_{\epsilon_O} \, y''$.
  \item Let the \emph{error} of a computer be the concatenation of tuples $\epsilon_{I}$ and $\epsilon_{O}$ characterizing the input and output error relations: the error is $\left(\epsilon_{i_1}, {\ldots}, \epsilon_{i_p}, \epsilon_{o_1}, {\ldots}, \epsilon_{o_q}\right) \in \mathbb{R}^{p + q}$.
  \item Crucially, imprecision during input and output notwithstanding, the user may still obtain (i.e.,\ measure) the \emph{correct} answer to the computation.
For example, if the computation being performed is expected to supply a natural-number output value, which is presented as the real-number value of some parameter of the system, then, by \emph{rounding} the real number to the nearest integer (so as to convert the \emph{measured} output value into the \emph{interpreted} output value; we use `$\iota$' to denote this `interpretation' mapping), the user can, \emph{granted sufficiently little imprecision}\footnote{In this natural-number example, ``sufficiently little imprecision'' amounts to the net effect of input imprecision, of the propagation/amplification thereof during computation, and of output imprecision rendering the measured output value distant by no more than $1/2$ from some correct output value, i.e.,\ an output value related via $\smile_\Phi$ to the intended input value.}, recover the correct answer; it is this ``sufficiently little imprecision'' that we formalize in our definition of the computational resource of \emph{precision}.
\end{itemize}

We summarize now the flow of computation in the above bullet points.
Without imprecision, we should have the ideal situation in which the user supplies (intended) input value $x$ to computer $\Phi$, and receives (measured) output value $y$ satisfying $x \smile_\Phi y$.
However, imprecision renders the flow as follows: \emph{intended input value} $x$ is conveyed to $\Phi$ imprecisely as \emph{implemented input value} $x'$ ($x \, R_{\epsilon_I} \, x'$), with which $\Phi$ computes, producing \emph{true output value} $y'$ ($x' \smile_\Phi y'$), which the user measures imprecisely as \emph{measured output value} $y''$ ($y' \, R_{\epsilon_O} \, y''$), which, in turn, the user interprets (based, typically, on the expected format of the output value; e.g.,\ by rounding or similar) as \emph{interpreted output value} $z := \iota\left(y''\right)$.

Intuitively, then, (intended) input value $x$ gives rise with strictly positive probability to (interpreted) output value $z$ if and only if there exist input value $x'$ and output values $y'$ and $y''$ such that $x \, R_{\epsilon_I} \, x' \smile_\Phi y' \, R_{\epsilon_O} \, y'' \stackrel{\iota}{\mapsto} z$.
In this case, we say that $x$ \emph{$\left(\Phi, \epsilon\right)$-yields} $z$, denoted $x \frown_{\Phi, \epsilon} z$, where $\epsilon = \left(\epsilon_I, \epsilon_O\right) \in \mathbb{R}^{p+q}$.

We are now in a position to define \emph{precision}.

\begin{definition}
Let $\Phi$ be a computer with input parameters $\left(i_1, V_{i_1}\right)$, {\ldots},\ $\left(i_p, V_{i_p}\right)$ and output parameters $\left(o_1, V_{o_1}\right)$, {\ldots},\ $\left(o_q, V_{o_q}\right)$, and let $x$ be an input value for $\Phi$.
\begin{itemize}
  \item An error $\epsilon \in \mathbb{R}^{p + q}$ is \emph{precise for $x$} if, for all $z$ such that $x \frown_{\Phi, \epsilon} z$, $x \smile_{\Phi} z$ (i.e.,\ imprecise though input/output may be, a correct output value is nonetheless produced, at least for input $x$: error $\epsilon$ is `sufficiently small' that it is corrected during interpretation).
  \item $\mathcal{E}_\Phi\left(x\right)$ denotes the set $\left\{\, \epsilon \in \mathbb{R}^{p + q} \;\middle\vert\; \epsilon \mbox{ is precise for } x \,\right\}$ of errors precise for $x$.
  \item $\mathcal{V}_\Phi\left(x\right) \in \left[0, \infty\right]$ is the $\left(p + q\right)$-dimensional Lebesgue measure of $\mathcal{E}_\Phi\left(x\right)$ (depending upon $p + q$, then, this is the length, area, volume or similar of $\mathcal{E}_\Phi\left(x\right)$).
  \item The \emph{precision required by $\Phi$ given $x$} is $P_\Phi\left(x\right) := \left\lfloor1/\mathcal{V}_\Phi\left(x\right)\right\rfloor \in \mathbb{N} \cup \left\{\infty\right\}$ (where, as usual, $1/0 := \infty$ and $1/\infty := 0$).
\end{itemize}
(Subscripts are often omitted when $\Phi$ is understood.)
\end{definition}

Intuitively, precision reflects the smallness of the set of errors corrigible by a computer, or, equivalently, the intricacy required (so as to achieve a corrigible error) of the computer's user when manipulating/measuring parameters.

As the notation suggests, $P$ is a \emph{computational} (commodity) \emph{resource} (recall Sect.~\ref{sec:intbaccpx}), on an equal footing with time $T$ and space $S$: just as one may ask how much \emph{time} will elapse during a computation, or how much \emph{space} is required in order for a computation to succeed, one may equally ask (numerically) how much \emph{precision} is required for computational success.
By considering this precision resource as a function of input size as in (\ref{eq:cpxfn}), then, one obtains the \emph{precision complexity} function
\[P^*\left(n\right) := \sup\left\{\, P\left(x\right) \;\middle\vert\; \left\vert x\right\vert = n \,\right\} \enspace .\]

Typically (though by no means necessarily), if input values $x_1$ and $x_2$ satisfy $\left|x_1\right| \leq \left|x_2\right|$, then $\mathbb{R}^{p + q} \supseteq \mathcal{E}\left(x_1\right) \supseteq \mathcal{E}\left(x_2\right) \supseteq \emptyset$, whence $0 \leq P\left(x_1\right) \leq P\left(x_2\right) \leq \infty$; $P^*$ is, in these typical cases, a non-decreasing function of input size. (We leave ``typical'' undefined, since we give here only an intuitive idea of the `direction' in which precision increases; cf.\ a Turing machine's typically, though not necessarily, requiring more time/space when processing larger input values.)

We defer additional detail regarding precision and precision complexity to (for example) \cite{thesis}.

\subsubsection{Analogue Factorization System}\label{sec:intbacafs}

By way of demonstration that there do indeed exist situations in which consideration not only of time and space, but also of precision, is necessary in order insightfully and correctly to analyse complexity, we now briefly outline a computing system (described in \cite{thesis,facana}, to which we defer further detail) that requires analysis so augmented. The system is an analogue computer that factorizes natural numbers in \emph{polynomial} time and space (cf.\ the exponential run-time required by the best publicly known digital algorithms), but that has \emph{exponential} precision complexity: the system cannot efficiently factorize large numbers, but the reason (namely, precision) for which it cannot is overlooked by standard, exclusively time-and-space complexity analyses; of such unconventional systems successful complexity analyses must heed correspondingly unconventional resources (e.g.,\ precision).

\paragraph{A geometric interpretation of factorization}

The chief observation behind the derivation of our system is that the problem of factorization---the search for proper divisors of a given natural number $n$---can be recast \emph{geometrically}.
One considers the hyperbola $y = n/x$: any point $\left(x, y\right)$ on the curve satisfies $xy = n$, and, in particular, any \emph{integer} point on the curve has integer coordinates $x$ and $y$ satisfying $xy = n$; such coordinates are \emph{factors} of $n$.
These integer points of interest (from the coordinates of which can be derived factors of $n$) are the points of intersection between the grid $\mathbb{Z}^2$ of integer points and the curve $y = n/x$; however, since the curve is a conic section, the sought points can equally be expressed as the intersection between $\mathbb{Z}^2$ and a certain cone in $\mathbb{R}^3$.\footnote{Specifically, the cone consists of all lines that meet the central axis $\left\{\, \left(x, x, \sqrt{2n}\right) \;\middle\vert\; x \in \mathbb{R} \,\right\}$ at the point $\left(0, 0, \sqrt{2n}\right)$, and do so at an angle of $\pi/4$ of a radian.
Formally, the grid $\mathbb{Z}^2$ in this three-dimensional context is $\mathbb{Z}^2 \times \left\{0\right\} = \left\{\, \left(a, b, 0\right) \;\middle\vert\; a, b \in \mathbb{Z} \,\right\}$.}
So as to construct our analogue factorization system, we physically implement (certain finite subsets of) these two structures---grid and cone---in such a way that their intersection can be identified and hence factors found.

\paragraph{Implementing the integer grid}

\begin{figure}[htbp]
  \centering
  \includegraphics[height=47.5mm]{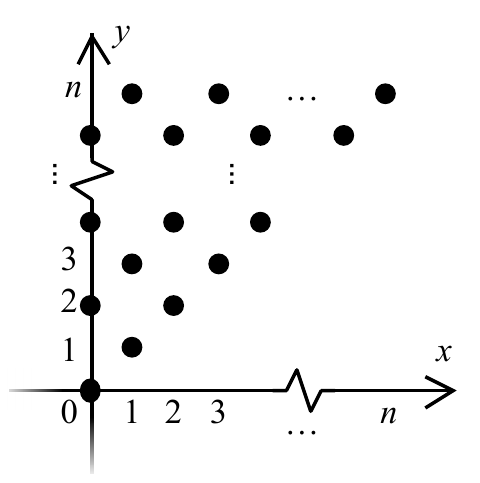}
  \caption{The $\frac{n^2 + 4n + 3}{4}$ points of grid $G_n$ (which lies in the plane $z = 0$).}\label{fig:gn}
\end{figure}

Of the grid $\mathbb{Z}^2 \times \left\{0\right\}$ we need implement only a finite part: the \mbox{$x$- and} $y$-coordinates of the sought points of intersection between the grid and the cone are factors of $n$, whence the subset $\left\{\, \left(a, b, 0\right) \in \mathbb{Z}^3 \;\middle\vert\; 1 \leq a, b \leq n \,\right\}$ suffices (for convenience, we consider instead $\left\{\, \left(a, b, 0\right) \in \mathbb{Z}^3 \;\middle\vert\; 0 \leq a, b \leq n \,\right\}$); furthermore, due to the symmetry of both cone and grid about the plane $y = x$ and to the commutativity of multiplication, identifying grid-cone-intersection point $\left(a, b, 0\right)$ yields the same factorization of $n$ (namely $n = ab$) as identifying $\left(b, a, 0\right)$, whence we need consider only $\left\{\, \left(a, b, 0\right) \in \mathbb{Z}^3 \;\middle\vert\; 0 \leq a \leq b \leq n \,\right\}$; finally, we assume that the value $n$ to be factorized is odd\footnote{Factorization of an even number may be achieved via repeated halvings (performed by Turing machine), of which the number is recorded, followed by factorization (performed by the analogue system described here) of the remaining, odd number. That the number of such halvings is necessarily merely logarithmic in $n$ gives that, asymptotically, the complexity of the system does not depend upon whether we include the Turing-machine halvings.}, whence all factors are odd, and so we need implement only points with both coordinates odd (for convenience, we implement the set
\[G_n := \left\{\, \left(a, b, 0\right) \in \mathbb{Z}^3 \;\middle\vert\; \begin{array}{cl}
           & 0 \leq a \leq b \leq n \\
    \wedge & a - b \mbox{ is even} \
  \end{array} \,\right\}\]
of points with \mbox{$x$- and} $y$-coordinates of the same parity---see Fig.~\ref{fig:gn}).

\begin{figure}[htbp]
  \centering
  \includegraphics[height=42mm]{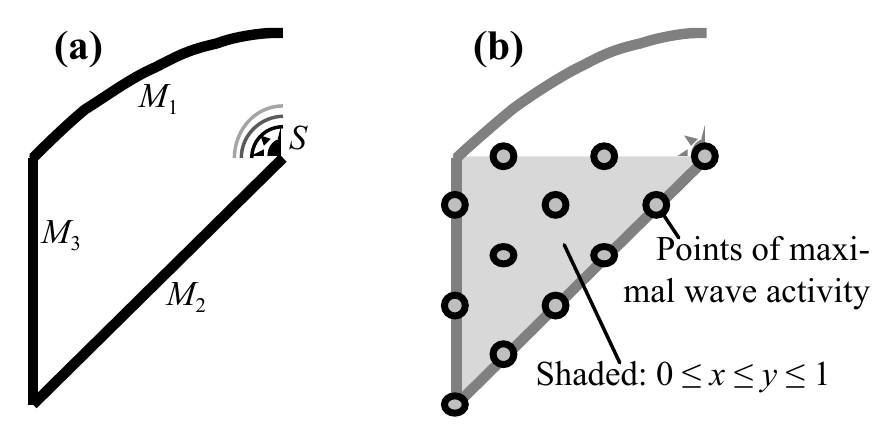}
  \caption{(a)~The source $S$ and mirrors $M_i$. (b)~The points of maximal wave activity, within the region $0 \leq x \leq y \leq 1$ (shaded), in the interference pattern produced by $S$ and $M_i$ (in this example, $n = 5$).}\label{fig:grid}
\end{figure}

We implement this grid $G_n$ by instantiating a certain interference pattern, of which the points of maximal wave activity model the elements of $G_n$.
The source that produces the pattern sits at $S := (1, 1, 0)$, has wavelength $\lambda := 2/n$, and is shielded so as to emit radiation only in the $x \leq 1 \leq y$ quadrant (and, so far as is practicable, only in the $z = 0$ plane); the radiation from this source is reflected by mirrors at $M_1 := \left\{\, \left(x, -x^2 / 2 + x + 1, 0\right) \in \mathbb{R}^3 \;\middle\vert\; 0 \leq x \leq 1 \,\right\}$, $M_2 := \left\{\, \left(x, x, 0\right) \in \mathbb{R}^3 \;\middle\vert\; 0 \leq x \leq 1 \,\right\}$ and $M_3 := \left\{\, \left(0, y, 0\right) \in \mathbb{R}^3 \;\middle\vert\; 0 \leq y \leq 1 \,\right\}$ (see Fig.~\ref{fig:grid}(a)).
The resultant interference pattern within the region $0 \leq x \leq y \leq 1$ (and $z = 0$) has maxima at precisely those points $\left(x, y, 0\right)$ where $nx$ and $ny$ are integers of the same parity (see Fig.~\ref{fig:grid}(b)); these maxima model the integer grid points of $G_n$ (where, explicitly, we interpret a point $\left(a, b, 0\right)$ of maximal wave activity as modelling $\left(na, nb, 0\right) \in G_n$)---compare Figs.~\ref{fig:gn} and \ref{fig:grid}(b).
(Thus, we employ a scaling of factor $1/n$ when implementing the system: a point $\left(x, y, z\right)$ in the mathematically abstract world of $G_n$, the curve $y = n/x$ ($z = 0$), etc.\ corresponds to the point $\left(x/n, y/n, z/n\right)$ in the `real world' inhabited by our apparatus---$S$, $M_i$, etc.)

\paragraph{Implementing the cone}

\begin{figure}[htbp]
  \centering
  \includegraphics[height=98mm]{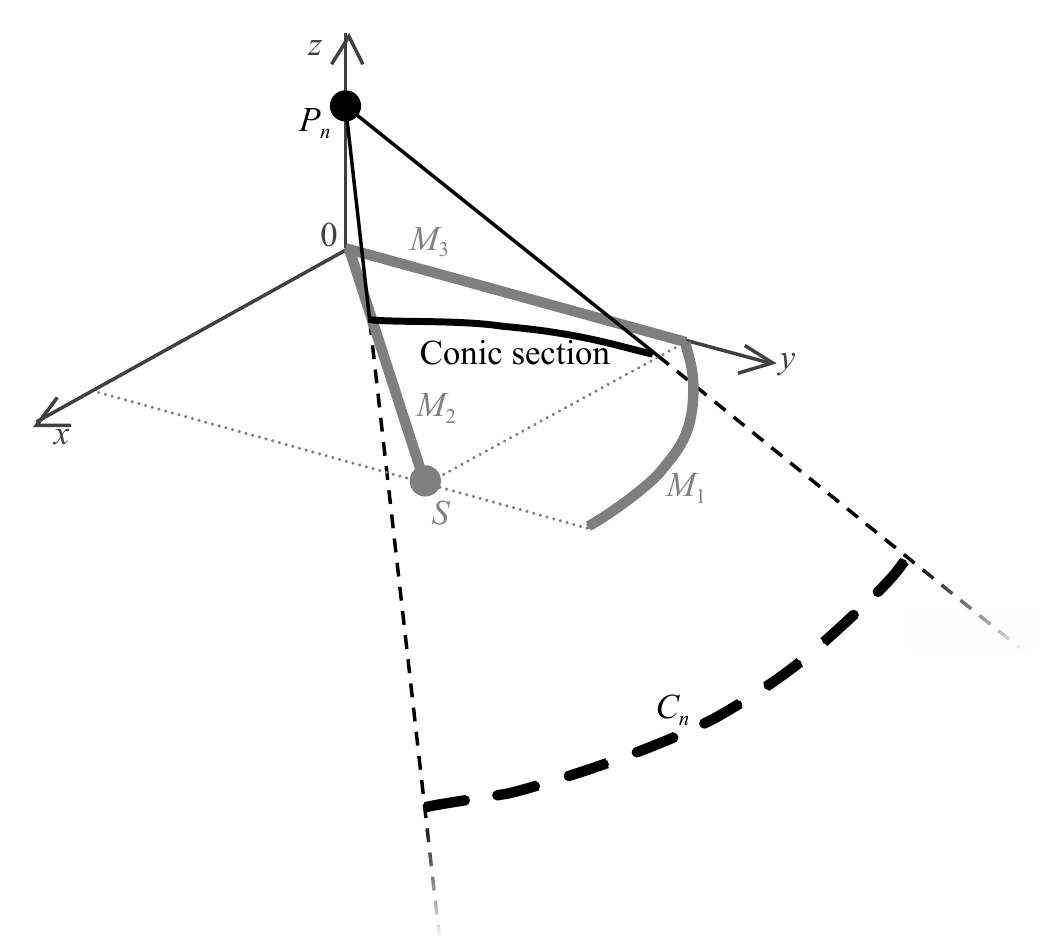}
  \caption{The apparatus implementing the cone: $P_n$ models the cone's \emph{vertex}, whilst $C_n$ occupies a circular arc in the \emph{surface} of the cone. The cone intersects the plane $z = 0$ containing the integer-point grid (produced by $S$ and $M_i$, which are shown in grey for context) along the conic section (pictured) that models the curve $y = n/x$.}\label{fig:cone}
\end{figure}

Having described the implementation of the grid $G_n$ of integer-coordinate points, we turn now to the implementation of the cone.
The vertex of the cone we model using a second source $P_n$ of waves, positioned at $\left(0, 0, \sqrt{2/n}\right)$; this, together with a sensor $C_n$ occupying a certain circular arc\footnote{The curve of $C_n$ is $\left\{\,\left(x, 2 - x, z\right) \in \mathbb{R}^3 \;\middle\vert\;
  \begin{array}{cl}
           & 2\left(x - 1\right)^2 + \left(z - \sqrt{2/n}\right)^2 = 2 \\
    \wedge & z \leq \frac{1 - n}{1 + n}\sqrt{2/n} \\
    \wedge & 2 - x \geq 1 \
  \end{array}
  \,\right\}$.}, uniquely specifies the cone, which is deemed to consist of those lines passing through both $P_n$ and a point on the circle containing $C_n$.
See Fig.~\ref{fig:cone}.

By construction, we have that the cone intersects the plane containing the grid of integer points (produced by $S$ and $M_i$) along the desired conic section $y = n/x$ (suitably transformed by the `abstraction-to-implementation modelling map' $\left(x, y, z\right) \mapsto \left(x/n, y/n, z/n\right)$ described above).

\mbox{}

\noindent Having implemented the two constituent parts of the analogue system---the grid of integer points and the cone\mbox{---,} we should like to be able to identify their \emph{intersection}, because, recall, the (\mbox{$x$- and} $y$-) coordinates of points belonging to both structures yield factors of $n$.
Provided that the two sources $S$ and $P_n$ produce waves of a `suitable' nature (and such choices of wave do exist---see Sect.~\ref{sec:intbacafs}~\emph{Example implementation} below), such identification is indeed possible: since the integer points in $G_n$ are modelled as points of maximal wave activity in the interference pattern produced by $S$ and $M_i$, it suffices that it is evident at sensor $C_n$ whether a ray passing from $P_n$ to $C_n$ (which necessarily passes through a point on the desired conic section $y = n/x$ in the grid's plane) has passed through a point of maximal wave activity; from the coordinates of points on $C_n$ at which this maximal-activity property is evident, one may then calculate (efficiently, via Turing machine) factors of $n$.

We now make these comments more concrete by describing an example implementation of the system; crucially, this entails our specifying an instance of the ``waves of a `suitable' nature'' alluded to above.

\paragraph{Example implementation}

\begin{figure}[htbp]
  \centering
  \includegraphics[height=89mm]{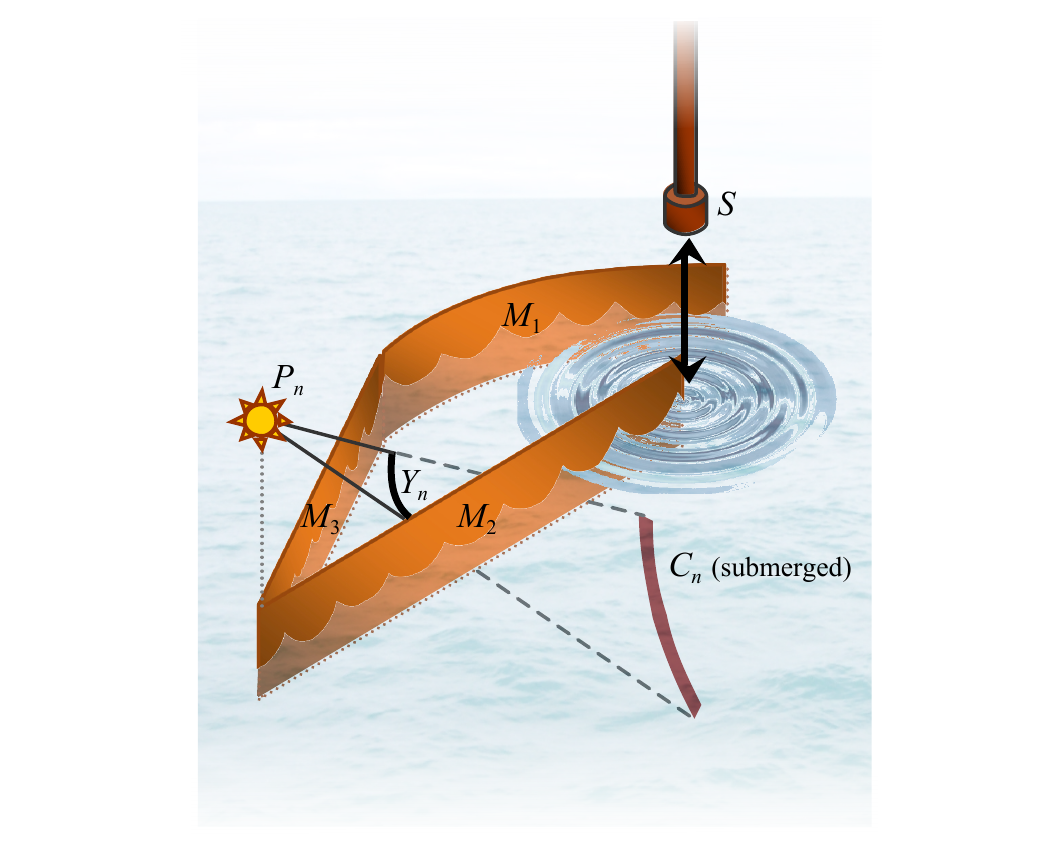}
  \caption{A water-wave ($S$, $M_i$) and visible-light ($P_n$, $C_n$) implementation of the analogue factorization device of Sect.~\ref{sec:intbacafs}.}\label{fig:water}
\end{figure}

Suppose that the interference pattern of $S$ and $M_i$ consists of \emph{water waves}: the source $S$ is a device that (sinusoidally, say) disturbs one point on the surface of a body of water (from which point ripples radiate), and the mirrors $M_i$ are reflective barriers protruding from the water---see Fig.~\ref{fig:water}.
This establishes our grid: the integer points of $G_n$ are modelled as points (within the triangular region of the water's surface between $M_2$ and $M_3$) of maximal wave activity, with calmer (i.e.,\ shorter-amplitude) water at the non-integer points between.

Suppose further that the second source $P_n$, suspended above the surface of the water, produces \emph{visible light}, which shines down into the water and, notably, onto the submerged \emph{light sensor} $C_n$ (which is suitably positioned so as to compensate for still-water refraction: in the absence of waves from $S$, light from $P_n$ shining through the curve---let us call it $Y_n$---modelling conic section $y = n/x$ in the water's surface would arrive at $C_n$, refraction notwithstanding); again, see Fig.~\ref{fig:water}.

Crucially, a ray of light from $P_n$ passing through a point $A$ on $Y_n$ on the water's surface (on which curve we wish to identify integer, i.e.,\ maximal-water-wave-activity, points) arrives at the corresponding point $B$ on $C_n$ with an intensity that betrays the amount of (water-) wave activity at $A$:\footnote{This is the essence of waves' being of a `suitable' nature: waves produced by $S$ must be of such a type that they affect those produced by $P_n$, as they propagate to $C_n$, in such a way that maximality or otherwise of those from $S$ may be determined at $C_n$.} if $A$ is a point of \emph{calm} water (which is the case at maximal distance from grid points), then light from $P_n$ arrives steadily at $C_n$, resulting in maximal brightness when summed over a (water-) wave cycle; if, instead, $A$ is an \emph{undulating} point on the water's surface, then light will reach $C_n$ only intermittently, due to the periodically fluctuating refraction at $A$.
If $A$ is \emph{maximally} undulating---if it is one of our sought points of maximal wave activity\mbox{---,} then this is evident as a \emph{minimal} amount of light (summed over a water-wave cycle) from $P_n$ reaching $B$.
The user of the system, then, employs $C_n$ in order to retrieve the coordinates of a minimally lit point $B$, from which coordinates may be calculated the coordinates of the corresponding surface point $A$; we reiterate that, (a)~since $A$ is a point through which, in the absence of water-waves from $S$, light from $P_n$ would pass to $C_n$, it is on the curve $Y_n$ that models $y = n/x$, and that, (b)~by minimality of light at $B$ and by the comments of this paragraph, $A$ is a point of maximal wave activity---therefore, $A$ is a sought grid point on $Y_n$, and its coordinates (which may be calculated from those of $B$) yield factors of $n$.

\paragraph{Using the system}

We summarize now the process whereby the analogue system (in its general form, not necessarily the water/light embodiment of Sect.~\ref{sec:intbacafs}~\emph{Example implementation}) yields a factor of $n$.
\begin{enumerate}
  \item\label{pt:tmip} First, given the value $n \in \mathbb{N}$ to be factorized, the user computes two values---$2/n$ and $\sqrt{2/n}$. This computation may be performed via standard, Turing-machine-style computation.
  \item\label{pt:setip} Secondly, these values are supplied, \emph{most probably imprecisely}, to the input parameters of the analogue device (recall the discussion in Sect.~\ref{sec:intbacpre} of input/output parameters): the wavelength of $S$ is set to $2/n$, and the height (i.e.,\ $z$-coordinate) both of $P_n$ and of the centre of the circle containing $C_n$ to $\sqrt{2/n}$.
  \item\label{pt:proc} Having so set the input parameters, the analogue computation takes place via propagation of waves from $S$ and $P_n$; notably, an interference pattern (of radiation from $P_n$ of varying intensity) is formed along sensor $C_n$.
  \item\label{pt:getop} From readings of sensor $C_n$, the user determines (again, \emph{most probably imprecisely}) the $x$-coordinate ($c$, say) of a minimally lit point on this sensor.
  \item\label{pt:tmop} Finally, from this coordinate $c$, the user computes (again via Turing machine or similar) the value $\sqrt{\frac{nc}{2 - c}}$, which is, we claim (with informal justification implicit above and formal deferred to \cite{thesis}), a factor of $n$.
\end{enumerate}

\paragraph{Time and space complexity}

We claim (and defer to \cite{thesis} justification beyond that offered by the comments here) that the analogue system---consisting of steps~\ref{pt:setip} to \ref{pt:getop} in Sect.~\ref{sec:intbacafs}~\emph{Using the system}---requires space and time merely \emph{constant} in the size of the value $n$ being factorized: space because the analogue apparatus is, for all $n$, contained within the constant-size, $n$-independent cuboid $\left[0, 2\right] \times \left[0, 2\right] \times \left[-\sqrt{2}, 2\sqrt{2}\right]$; time because the user need wait only for waves (the propagation velocity of which we suppose to be independent of wavelength) to propagate over a constant, $n$-independent distance within this cuboid.

To this constant time/space overhead, we add the \emph{polynomial} (specifically, \emph{quadratic}) cost incurred during the Turing-machine calculations of steps~\ref{pt:tmip} and \ref{pt:tmop}.
(Input values $2/n$ and $\sqrt{2/n}$ need be calculated only with sufficient precision to allow retrieval of $n$ \emph{given that $n \in \mathbb{N}$}; output value $\sqrt{\frac{nc}{2 - c}}$ need be found only to the \emph{nearest integer}; this results in the quadratic complexity claimed---see \cite{thesis} for further detail.)

As a whole, then, the factorization system of Sect.~\ref{sec:intbacafs} has polynomial time and space complexity; this represents a marked improvement over the exponential time complexity of the most efficient publicly known \emph{digital-computer} factorization methods.
The catch, as one may imagine given the discussion of Sect.~\ref{sec:intbacpre}, is the \emph{precision complexity} of the system, which, we see below, is exponential.

\paragraph{Precision complexity}

We consider the input parameter $\lambda$, the wavelength of source $S$.\footnote{One may treat other input/output parameters analogously. However, we show below that parameter $\lambda$'s contribution to the system's precision complexity is exponential, whence the overall precision complexity is itself exponential, regardless of the contributions of other parameters; accordingly, such parameters are not explicitly considered.}
The user's intention, recall, it to set $\lambda$ to the value $2/n$, where $n \in \mathbb{N}$ is the number to be factorized.
Let us suppose
\begin{itemize}
  \item that the input to $\lambda$ of value $2/n$ suffers an \emph{additive error} characterized by error term $\epsilon \geq 0$: the value actually supplied to $\lambda$ is an arbitrary element of $\left[2/n - \epsilon, 2/n + \epsilon\right]$; but
  \item that the system performs \emph{error correction} based upon the fact that $2/\lambda$ is expected to be a natural number: given wavelength $\lambda = 2/\nu$ for arbitrary $\nu \in \mathbb{R}^+$, the system acts as though the number to be factorized is the nearest integer $\left\lfloor\nu + 1/2\right\rfloor$ to $\nu$.
\end{itemize}

The purpose of the resource of precision is to quantify the fidelity with which $\lambda$ must be set (and the other parameters set/measured) in order that the computation be performed correctly---to quantify, that is, how small $\epsilon$ (and the other parameters' analogous error terms) must be in order that the received, error-corrected input value $\left\lfloor\nu + 1/2\right\rfloor$ to be factorized coincide with $n$ (and, when considering also output parameters, that the measured and interpreted output value coincide with a factor of $n$).

$\epsilon$ is sufficiently small in this sense if and only if the supplied wavelength $2/\nu \in \left[2/n - \epsilon, 2/n + \epsilon\right]$ necessarily falls in the interval $\left(\frac{2}{n + 1/2}, \frac{2}{n - 1/2}\right]$ of values that the system corrects to $2/n$, which is the case if and only if $\left[2/n - \epsilon, 2/n + \epsilon\right] \subseteq \left(\frac{2}{n + 1/2}, \frac{2}{n - 1/2}\right]$, which, in turn, is the case if and only if $\epsilon < \frac{1}{n\left(n + 1/2\right)}$.
Thus, the set of corrigible errors $\epsilon$ in input parameter $\lambda$ is $\left[0, \frac{1}{n\left(n + 1/2\right)}\right)$, whence $\lambda$ contributes to the system's overall precision complexity a multiplicative factor of $n\left(n + 1/2\right)$, which increases quadratically with $n$ and therefore \emph{exponentially} with the size of $n$.
Thus, regardless of the contributions from other input/output parameters, the system's precision complexity is exponential.

We state again that the system does not offer efficient means of factorizing large numbers, though not because of its time or space complexity, but rather because of its \emph{precision} complexity, which is seen in the above discussion quantitatively to impose an exponential cost.
We describe now a framework in which may be made complexity analyses suitable for such situations (analyses, that is, that heed not only time and space but also precision and more).

\subsubsection{Computational-Model-Independent Framework of Complexity Theory}\label{sec:intbacfra}

In \cite{thesis}, we develop a framework in which one may insightfully analyse and meaningfully compare the complexity of instances of many different computational paradigms, with respect to many different resources.
We recap now the salient aspects of the framework.

\paragraph{Resource generality}

The crucial observation from Sect.~\ref{sec:intbacafs} is that, in the specific case of the analogue factorization system, an insightful complexity analysis must heed not only the standard resources of time and space but also the non-standard resource of \emph{precision} (for else one overlooks the true, exponential complexity of the system).
More generally, the observation is that, when analysing the complexity of unconventional (analogue, chemical, optical, quantum, {\ldots})\ computers, the analysis must be with respect to accordingly unconventional resources (precision, energy, etc.,\ in addition to time and space).

Accordingly, the first step in the implementation of our complexity framework is a generalization of \emph{resource}, such that one considers not merely time and space, nor yet merely time, space and precision, but rather \emph{arbitrary resources} (in fact, this broad conception of resource is to be refined---see Sect.~\ref{sec:intbacfra}~\emph{Normalization} below; arbitrary resource offers a suitable starting point, however).

\paragraph{Overall complexity}

Having introduced many different (in fact, \emph{arbitrary}) resources, it is no longer clear how to define the \emph{overall} complexity of a computation.\footnote{A measure of overall complexity, furthermore, is highly desirable, both \emph{practically}---by comparing with respect to the $\mathcal{O}$-preorder of complexity functions the respective overall complexity of two systems, one ascertains which is the more efficient and, therefore, preferable to use---and \emph{theoretically}---the ability to measure and compare systems' overall complexity allows identification of the most efficient (known) system for a particular computational task, and, therefore, of the tightest (available) upper bound on the complexity of the task itself.}
When one considers Turing machines and similar, the task is unproblematic: such systems consume only the resources of time and space, and, for any computation, the consumption of the former is an upper bound for that of the latter (since writing to memory takes time---more precisely, each time-step sees the computer write to at most one tape cell), whence time complexity offers an adequate measure of overall complexity (thus, the (overall) more efficient Turing machine is precisely the asymptotically faster).
However, when considering systems conforming to more exotic (analogue, chemical, quantum, {\ldots})\ computational paradigms, it is certainly not clear which (if any) of the many complexity functions, corresponding to the many resources consumed, successfully captures overall complexity.

Intuitively, the task is to ascertain which of these many resources are `relevant' to a computation---which, asymptotically, are consumed in sufficiently significant quantity that they should contribute to any reasonable measure of overall complexity.
We capture this criterion of relevance by defining \emph{dominance}: a resource is deemed \emph{dominant} for a computation if its complexity function is maximal in the preorder $\lesssim$, where, for complexity functions $f$ and $g$, `$f \lesssim g$' is defined to mean `$f \in \mathcal{O}\left(g\right)$' (we reiterate that further detail is given in \cite{thesis}).

A measure of \emph{overall complexity} of a computation may then be defined to be the sum of the complexity functions corresponding to those resources that are dominant for the computation (this definition is consistent with the observation above that, for a Turing machine---for which time is necessarily dominant\mbox{---,} time complexity and overall complexity are one and the same).

\paragraph{Normalization}

As is clear from Sect.~\ref{sec:intbacfra}~\emph{Overall complexity}, for dominance usefully to serve its intended purpose it must determine definitively which of a collection of resources are `relevant' to a computation and so which contribute to the overall complexity.
However, without constraining what is acceptable as a valid resource (recall that, in Sect.~\ref{sec:intbacfra}~\emph{Resource generality} above, no restriction is made), dominance unfortunately does not have this property; we illustrate this now.

Recall from Sect.~\ref{sec:intbaccpx} the two standard resources consumed by a Turing machine: $S$ is the number of distinct tape cells to which are written during a computation, $T$ the number of time-steps that elapse.
Since, for any input value $x$, $T\left(x\right) \geq S\left(x\right)$, we have that time is dominant: time is deemed to be no less relevant to the computation than space; this is exactly what one expects and hopes of dominance.
However, one may measure space not with $S$ but with, say, $S'$, given by $S'\left(x\right) := 2^{S\left(x\right)}$---we merely \emph{relabel} quantities of space: rather than counting them as $0$, $1$, $2$, $3$, {\ldots},\ we (unusually but validly) use $1$, $2$, $4$, $8$, {\ldots}
It is then perfectly possible, our having artificially exaggerated the relevance of space by using $S'$ rather than $S$, that $S'$ (not $T$) becomes dominant. Whereas we should like that dominance determines definitively which of time and space is the more relevant, in fact we may \emph{engineer} which is deemed more relevant simply by altering the way in which quantities are measured or labelled.

So as to remove this undesirable ability to engineer, we stipulate that any resources considered must be \emph{normal}.
A normal resource, roughly, is one that can attain any natural-number value.
For example, $S$ is normal since, for any natural number $a$, there exist a Turing machine $M$ and an input value $x$ such that $S_M\left(x\right) = a$; $S'$ is not normal since, when $a$ is not a power of two, for no pair $\left(M, x\right)$ do we have that $S'_M\left(x\right) = a$.
\emph{Normalization} is a process whereby an arbitrary, unrestricted resource is converted into an order-isomorphic \emph{and, crucially, normal} resource; e.g.,\ as one may imagine, $S'$ normalizes to $S$.

Once we stipulate that resources be normal, then, one may no longer engineer as above which of time and space is more relevant, for he is no longer free to consider the (abnormal) resource $S'$; one must instead quantify space using $S$, and is consequently led by dominance to the irrevocable conclusion that time is more relevant than space.\footnote{We stress that this conclusion holds in the Turing-machine example presented above, though not necessarily more generally for other computational paradigms.}

See \cite{thesis,bbwr} for more detail concerning normalization.

\paragraph{Summary}

A much fuller account is given in \cite{thesis}, but we outline above the main features---\emph{resource generality}, \emph{dominance} and the corresponding notion of \emph{overall complexity}, and \emph{normalization}---of our model-independent framework of computational complexity theory.
The framework allows analysis and comparison of the complexity of computers conforming to many paradigms and consuming many resources.

\subsection{Motivation}\label{sec:intmot}

We note from the preceding discussion that there exist many (commodity) resources: the traditional \emph{time} and \emph{space}, as well as the non-standard \emph{precision} (which we define above by way of illustration), \emph{energy}, \emph{mass}, \emph{thermodynamic cost}, \emph{material cost}, and many more besides; see \cite{ucuc, thesis} for further detail.
We recall from (\ref{eq:cpxfn}) that, from these many resources, one derives in a uniform way the corresponding complexity functions; if one observes that his non-standard computer consumes some new, non-standard resource, then he acquires also a new complexity function.
To these many complexity functions one may apply various complexity-theoretic techniques and tools in order to analyse and compare computers' efficiency and computational tasks' difficulty; Sects.~\ref{sec:intbacafs}~\emph{Precision complexity} and \ref{sec:intbacfra} above give a feel for such application, whilst \cite{thesis} furnishes additional detail on this topic.

It is desirable, we suggest, to employ these complexity-theoretic techniques and tools in the analysis not only of computational processes (as described above), but also of \emph{cryptographic protocols} and similar; this suggestion motivates Sect.~\ref{sec:crs} of the present paper.
More explicitly, if one were able to abstract from a cryptographic protocol entities that behave (in some appropriate sense) as \emph{resources}, then there would result via (\ref{eq:cpxfn}) entities that behave as \emph{complexity functions}, to which could be applied the existing and understood arsenal of complexity-theoretic techniques so as to analyse the protocol.

The chief intuition here is that we wish to enable the formulation as a resource of the \emph{security} (whatever that may mean) of a cryptographic protocol.
This having been done, (\ref{eq:cpxfn}) would then define the protocol's \emph{security complexity}, about which one may reason via complexity-theoretic means with a view to proving/disproving\footnote{Implicit in this talk of proof/disproof are the twin motivations of \emph{information assurance} and \emph{signals intelligence}. The latter requires that we define the notions of the present work in such a way that one is able not only to identify the presence of insecurity in others' systems, but also to determine and exploit the root causes of such insecurity.} that the protocol is secure.

Hence, we advocate here (though defer implementation and investigation largely to future work) a \emph{resource-centric framework} in which many aspects of cryptographic protocols may be captured as resources so as to allow security analysis of the protocols via previously inapplicable, complexity-theoretic techniques. Such a framework addresses the author's personal interpretation of and thoughts on this special issue's topic, `information security as a resource'.

\section{Cryptographic Resources}\label{sec:crs}

So as to illustrate the abstraction of resource-like entities from cryptographic protocols, we consider the example of \emph{pubic-key cryptography}.
The intent here is for sender Alice to transmit a message to recipient Bob (the two are spatially distant), without eavesdropper Eve's being able to obtain this message (Eve, we assume, intercepts all communications between Alice and Bob and does so undetected); that we consider specifically \emph{public-key} cryptography implies that Alice and Bob do not, prior to the protocol's commencing, agree upon any shared information (a key, for example) of which Eve is unaware.

The scheme's outline is as follows.
\begin{itemize}
  \item Bob generates two keys: one private (that Bob does not at any point transmit) and one public.
  \item Bob sends the public key to Alice (and also, we must assume, to Eve).
  \item Alice encrypts the message, with the encryption process taking as a parameter Bob's public key.
  \item Alice sends the encrypted message to Bob (and to Eve).
  \item Bob decrypts the message, with the decryption process taking as a parameter his private key; this yields Alice's original message.
\end{itemize}

During the protocol, then, Eve gains access to Bob's public key and the encrypted message.
Hence, for the protocol to function as desired, it must be the case not only (a)~that the relationship between Bob's public and private keys is such that his decryption process does indeed (efficiently) yield the original message, but also (b)~that any `cheat' computation (that Eve might perform) that takes as input the public key and the encrypted message, and that returns the original message, is computationally infeasible; i.e.,\ (a)~Bob can successfully and efficiently retrieve Alice's communication, whereas (b)~Eve cannot.

If for example the specific implementation of public-key cryptography under consideration is \emph{RSA} \cite{rsa}\footnote{Perhaps a fairer name for the cryptosystem than `RSA' would be `C', for, whereas Rivest, Shamir and Adleman \emph{publicly} introduced the system in 1978, an equivalent form had in fact been discovered (unannounced) in 1973 by Clifford Cocks.}, then such a `cheat' computation whereby Eve can retrieve the original message is \emph{natural-number factorization}: finding the factors of the public key reveals the private key, given which Eve could decrypt using the same process as Bob.
Fortunately for Alice and Bob, although the difficulty of factorization has not been rigorously established (the equivalent decision problem is believed not to be $\ccn{NP}$-hard, moreover), neither is there publicly known an efficient and practicable means of factorization (attempts spanning many mathematician-millennia and utilizing both standard and non-standard computation notwithstanding).

This suggests that, at least for as long as factorization remains difficult in practice, RSA offers a secure public-key scheme.
Implicit in this suggestion, however, is the assumption that a \emph{complexity-theoretic} consideration (of the difficulty or otherwise of Alice's, Bob's and Eve's respective processes) captures all aspects of the protocol relevant to an analysis of its security---that potential insecurities of the protocol are necessarily \emph{computational} in this sense; this overlooks the possibility that, the apparent difficulty of factorization notwithstanding, Eve may be able to exploit for example \emph{side-channel information} such as the time taken for Alice to encrypt the message or the amount of memory used by Bob in generating the public and private keys; if such information were to betray to Eve knowledge of the message itself or of Bob's private key, then there may exist within the protocol insecurities not predicated upon efficient means of factorization.\footnote{Note that we mean not to suggest that such side-channels can necessarily be exploited in the specific case of RSA. Rather, we mention RSA as an illustrative public-key system with which the reader is likely to be familiar, and (separately) introduce side-channels as an aspect that we wish to include in our analyses of \emph{arbitrary} cryptographic protocols.}

Accordingly, we advocate in the present paper an approach to the analysis of cryptographic systems that heeds not only the computational aspects of the protocols so analysed, but also the non-computational: those relating to communication, information, cryptographic primitives, and so on.
More precisely, we advocate the formulation of these aspects (both computational and non-) as \emph{resources}, which can then be analysed using the existing tools of complexity theory.
We discuss in more detail now each of these categories of aspects/resources.

\subsection{Computation}\label{sec:crscmp}

The capture as resources of \emph{computational} aspects of a protocol or scheme requires no new machinery: one can apply the existing techniques of standard complexity theory (\cite{papa}) and its model-independent generalization (\cite{thesis} and Sect.~\ref{sec:intbacfra}) to any computations that take place during either the protocol or potential attacks thereof.
In the case of public-key cryptography, for example, these computations include Bob's key generation, Alice's encryption, Bob's decryption (all of which one hopes are easy) and any computation (which one wants to be difficult) whereby Eve may obtain the message.

By so analysing these computations, one quantifies computational resources such as time and space, and, if the protocol involves unconventional computers, possibly also precision, energy, etc.

\subsection{Communication}

We wish also to formulate as resources the \emph{communication} aspects of protocols.
In public-key cryptography, for example, there is communication (of public keys and encrypted messages) between Alice and Bob, and also, unintentionally, between them and Eve.
We should, therefore, like to accommodate in our analyses of protocols such communication-theoretic resources as \emph{channel capacity}.

\subsection{Information}

A further category of resource that we wish to capture concerns the \emph{information-related} aspects of protocols.
Again considering the example of a public-key cryptographic system, there are present various items of information: the message (both plain and encrypted), the public and private keys, and, crucially though less obviously, \emph{side-channel} information.

It may for example be the case that, if Eve were able to monitor the time taken by Alice in encrypting the message, then that duration would betray to Eve some significant information about the message itself; alternatively, it may be that, if Eve were able somehow to measure the memory usage of Bob's key-generation routine, then that knowledge would betray to Eve something of the nature of Bob's private key (whence extracting the key itself may become tractable).\footnote{\label{fn:kwies}We note that the difficulty in exploiting a side-channel is typically in identifying that the channel exists, rather than in using it once it has been discovered. Consequently, in order successfully to accommodate side-channels and related phenomena within our framework, we must consider not only the `commodity' resources (run-time, memory space, etc.)\ consumed \emph{during} Eve's use of a side-channel, but also the `manufacturing' costs that Eve incurs \emph{initially} whilst discovering/contriving the channel. See \cite{ucuc} for discussion of commodity and manufacturing resources.}

Although these potential betrayals of information are not explicit in the description of the protocol (they are not explicitly represented by communication channels or similar), they nonetheless constitute an implicit flow of information for which we should like to account when assessing (via the resource-centric approach advocated here) the security of the protocol.

\subsection{Cryptographic Primitives}\label{sec:crsprm}

We comment now on the generality and applicability of the resource-centric framework described in the present paper.

Note that we allude above to concepts such as ease/difficulty for Alice, Bob and Eve; this of course assumes that these three named roles exist within the protocol (and attacks thereof) under consideration. Whereas these roles do indeed exist in standard cryptographic (e.g.,\ public-key) protocols, they may not be present in other protocols (such as \emph{coin-tossing schemes}) that nonetheless feature similar issues of information sharing, mutual distrust, the availability of zero-knowledge proofs, etc.,\ and that we should nonetheless like to accommodate (alongside the traditional Alice-Bob-Eve cryptosystems) in our resource-centric framework.

Accordingly, so as to accommodate schemes and protocols that do not feature Alice, Bob and Eve (or, more generally, that do not feature `goodies' and `baddies'), it is desirable to conduct analyses focussing not on named agents, but rather on the \emph{abilities and inabilities} of (anonymous) agents (whence, if we are in fact analysing an Alice-Bob-Eve set-up, we may \emph{derive} which agent is which).

Consequently, we wish to consider (in the context of our resource-centric approach) the \emph{cryptographic primitives} (such as one-way functions, trapdoor functions and pseudorandom number generators) that may or may not be available to (anonymous) agents, and, furthermore, to place these primitives on an equal footing with other (computation-, \mbox{communication- and} information-related) resources, such that trades-off between primitives and other resources can be considered.\footnote{Note that an excellent starting point for our introducing primitives into our resource-centric framework is the approach of Ran Canetti; see, for example, \cite{can1,can2}.}

\subsection{Summary}

The purpose of the resource-centric framework advocated in the present paper is to allow insightful analysis of protocols (cryptographic systems, coin-tossing schemes, etc.)\ with respect to many aspects: computation, communication, information (including, for example, side-channel information and the existence of subliminal channels), availability to the respective agents of primitives, etc.

Intuitively, the more such aspects that are considered, the greater the likelihood of identifying any insecurities present in the protocol (or else the greater the significance of lack of such identification).
If, for example, one were to consider only the \emph{computational} aspects of RSA, then he might convince himself that the system is secure for as long as factorization is difficult; however, this may overlook insecurities (e.g.,\ side-channels) relating to features of RSA that are not inherently computational.

Accordingly, we wish to accommodate in our framework resources belonging to the various categories described in Sects.~\ref{sec:crscmp}~--~\ref{sec:crsprm}: computation, communication, information and primitives.
Ultimately, the aim is to define (in terms of its relationship with these resources) the resource of \emph{security}, about which we may then reason complexity-theoretically.\footnote{\label{fn:ofaw}Furthermore, having considered within the framework these diverse categories of resource---computation, communication, etc.\mbox{---,}\ it may be possible to establish the security of a protocol relative to assumptions that are not necessarily complexity-theoretic (`factorization is difficult' or similar) or physical (`Eve's intercepting a quantum channel is detectable by Alice and Bob' or similar), say, but that are taken from a more general class of assumptions concerning the many different categories of resource that we consider.}

\subsection{Security}

We consider now the form that the resource of security may take.
Prima facie, a reasonable first attempt at a definition would render security a one-dimensional quantity (a real number, say)---that depends upon key-size or similar (we once again have in mind public-key cryptography as our paradigmatic and illustrative protocol)---that is \emph{large} when Bob can easily obtain Alice's message but Eve cannot, and \emph{small} otherwise.
However, as is implicit in the phrases ``can easily''/``cannot [easily]'', this approach captures essentially nothing more than the standard computational-complexity view of the protocol: this measure of security reflects merely the complexity of Eve's `cheat' computation (e.g.,\ public-key factorization in the case of RSA), whereas, recall, we should like also to capture many non-computational aspects (the presence of side-channels, etc.).
It seems, then, that security should be a \emph{multi-dimensional} quantity that reflects not only the protocol's constituent computations but also its communication, information, primitives, etc.

Note that the execution of each of the sub-processes making up a protocol can incur costs in terms of resources belonging to these various categories: computation, communication, information, primitives, {\ldots}
For example, Alice's \emph{encryption} routine incurs a (hopefully small) computational cost, but no communication cost since encryption takes place locally on Alice's computer; on the other hand, the \emph{transmission} from Alice to Bob of the encrypted message incurs a communication cost (requiring a channel of a certain capacity for a certain amount of time), but no computational cost since, by definition of the sub-process under consideration, the message is encrypted and ready for transmission.
Incurred by each of these sub-processes, then, is a cost in only one category (computation for the former, communication for the latter).

\emph{If} this property of incurring cost in but a single category were to hold for all sub-processes and for all features of a protocol that one may consider, then there would be no apparent interaction between the respective categories, suggesting that one may consider each in isolation: analysis of the security of a protocol would then decompose into a series of separate searches for insecurities relating respectively to computation, communication, information and so on; the protocol would be deemed secure if and only if it were to pass the computation security test, and (separately) the communication security test, and (separately) the information security test, etc.
Note that each such test is already extensively studied and has its own associated literature (of which some is cited in the present paper); such decomposition would result in our framework offering nothing new.

However, if there exist features or sub-processes (we suggest, furthermore, that such do indeed exist) that fall into strictly more than one category of resource, if there arise trades-off between resources and/or primitives from different categories---if, in short, the different categories \emph{interact}\mbox{---,} then one is no longer free to view security analyses within the framework as series of non-interacting tests, having rather to consider the framework as a coherent whole: the successful analysis of a protocol's security requires \emph{simultaneous} consideration of the various facets represented by the different categories of resource/primitive.
We suggest moreover that the resource of \emph{security} does indeed straddle several categories, having non-trivial relationships with and dependencies on many different resources and primitives; hence, our analysis of protocols cannot, without sacrificing useful---even crucial---information, be decomposed, but must be performed within a coherent, all-encompassing framework.
It is the implementation, study and use of precisely this framework that we advocate here.

\subsection*{Acknowledgements}

We thank Mike Mislove for his help in developing the project of which the (proposed) work of Sect.~\ref{sec:crs} of this paper represents part.
We thank Omar Fawzi for discussions leading to the observation of Footnote~\ref{fn:ofaw}, Karoline Wiesner for discussions leading to the observation of Footnote~\ref{fn:kwies}, and Andreas Winter for general discussion about the ideas behind Sect.~\ref{sec:crs}.
We thank the \emph{Information and Computation} referees for their helpful comments and suggestions about this paper.
We acknowledge the generous financial support of the EPSRC (part of this work was funded by grant EP/G003017/1), and of the Leverhulme Trust and European Commission (which fund the author's current position).

\bibliographystyle{plain}
\bibliography{blakey.bib}

\end{document}